\documentclass{article}
\usepackage{waspaa21,amsmath,graphicx,url,times}
\usepackage{amsmath,graphicx}
\usepackage{tabularx}
\usepackage{cite}
\graphicspath{{Figs/}}
\usepackage{float}
\usepackage{graphicx}
\usepackage{epsf}
\usepackage{array}
\usepackage{makecell} 
\usepackage{amssymb}
\usepackage{lineno}
\usepackage{fixmath}
\def\<#1>{\mathinner{\langle#1\rangle}}
\usepackage{float}
\usepackage{caption}
\usepackage{subcaption}
\usepackage{tikz}
\usepackage{color}
\newcolumntype{P}[1]{>{\centering\arraybackslash}p{#1}}
\newcolumntype{Y}{>{\centering\arraybackslash}X}
\newcommand{\heading}[1]{\multicolumn{1}{c}{#1}}
\newsavebox{\foobox}

\title{Objective Metrics to Evaluate Residual-Echo Suppression \\ During Double-Talk}
\name{Amir Ivry \qquad Israel Cohen \qquad Baruch Berdugo \thanks{This research was supported by the Pazy Research Foundation and the ISF-NSFC joint research program (grant No. 2514/17). The authors thank Stem Audio for providing equipment and technical guidance.}}
\address{Andrew and Erna Viterbi Faculty of Electrical and Computer Engineering \\ Technion -- Israel Institute of Technology, Technion City, Haifa 3200003, Israel}

\begin{document}

\ninept
\maketitle

\begin{sloppy} 

\begin{abstract}
Human subjective evaluation is optimal to assess speech quality for human perception. The recently introduced deep noise suppression mean opinion score (DNSMOS) metric was shown to estimate human ratings with great accuracy. The signal-to-distortion ratio (SDR) metric is widely used to evaluate residual-echo suppression (RES) systems by estimating speech quality during double-talk. However, since the SDR is affected by both speech distortion and residual-echo presence, it does not correlate well with human ratings according to the DNSMOS. To address that, we introduce two objective metrics to separately quantify the desired-speech maintained level (DSML) and residual-echo suppression level (RESL) during double-talk. These metrics are evaluated using a deep learning-based RES-system with a tunable design parameter. Using 280 hours of real and simulated recordings, we show that the DSML and RESL correlate well with the DNSMOS with high generalization to various setups. Also, we empirically investigate the relation between tuning the RES-system design parameter and the DSML-RESL tradeoff it creates and offer a practical design scheme for dynamic system requirements. 
\end{abstract}


\section{Introduction}\label{sec:intro}
Hands-free communication often involves a conversation between two speakers located at near-end and far-end points. The near-end microphone can capture the desired-speech signal and two interfering signals: nonlinear echo produced by a loudspeaker playing the far-end signal, and background noises \cite{benesty1998better, benesty2001advances}. The acoustic coupling between the loudspeaker output and the microphone may lead to degraded speech intelligibility in the far-end due to echo presence \cite{sondhi1995stereophonic}. The most challenging scenarios are double-talk periods, when the desired speech and echo are captured by the microphone at the same time. To combat that, numerous nonlinear acoustic echo cancellation (NLAEC) systems were proposed to remove the nonlinear echo and to preserve the near-end speech \cite{guerin2003nonlinear,malik2012state,comminiello2013functional,halimeh2019neural,Ivry2021Nonlinear}. However, often there is still a mismatch between true and estimated echo paths, especially during the NLAEC convergence and re-convergence\cite{birkett1995limitations,mossi2010assessment}. As a result, the echo is not eliminated and the NLAEC should be followed by a residual-echo suppression (RES) system.

Human perception of speech quality is optimally evaluated using human subjective evaluation \cite{reddy2019scalable}. Lately, the objective deep noise suppression mean opinion score (DNSMOS) metric has been proposed to estimate human ratings and has shown great accuracy \cite{reddy2020dnsmos}. Regarding the task of RES, speech quality during double-talk is traditionally evaluated using the objective signal-to-distortion ratio (SDR) metric \cite{vincent2006performance}, e.g., in \cite{carbajal2018multiple,desiraju2019online,pfeifenberger2020nonlinear,chen2020nonlinear,fang2020robust, fang2020integrated}. Unfortunately, the SDR is affected by both desired-speech distortion and residual-echo presence, which renders it unreliable in predicting the DNSMOS and unreliable in predicting human perception of speech quality \cite{reddy2020dnsmos}.

This paper introduces two objective metrics that separately evaluate the desired-speech maintained level (DSML) and the residual-echo suppression level (RESL) during double-talk. Considering the RES system as a time-varying gain, the DSML is obtained by applying that gain to the desired speech and substituting the outcome in the definition of the SDR. The RESL is obtained by subtracting the desired speech from the double-talk segment and calculating the ratio of the noisy residual-echo before and after the gain is applied to it. To evaluate these metrics, we employ a deep learning-based RES system that also embeds a design parameter \cite{Ivry2021Deep}. Experiments are done with 280~h of real and simulated recordings in various scenarios and in high and low levels of echo and noise. Results show that the DSML and RESL have high correlation with human perception according to the DNSMOS, and high generalization to various setups, which renders them more suitable for speech quality evaluation than the SDR. We further investigate the empirical relation between tuning the design parameter and the DSML-RESL tradeoff it creates. Based on this relation, we offer a practical scheme for tuning the design parameter during training to optimally cope with dynamic system requirements.

The remainder of this paper is organized as follows. In Section~\ref{ProblemFormulation}, we formulate the problem. In Section \ref{sec:methods}, we introduce the DSML and RESL metrics.  Section~\ref{sec:system} covers the employed RES system and its tunable design parameter. Section~\ref{sec:setup} describes the database and additional performance metrics, and experimental results are presented in Section~\ref{sec:results}. Section~\ref{sec:conclusion} concludes this study.

\section{Problem Formulation}\label{ProblemFormulation}
Figure~\ref{fig:setup} depicts the RES scenario. Let $s\left(n\right)$ be the desired near-end speech signal and let $x\left(n\right)$ be the far-end speech signal. The near-end microphone signal $m\left(n\right)$ is given by
\begin{align}
m\left(n\right) = s\left(n\right) + y\left(n\right) + w\left(n\right), 
\label{eq:micModel}
\end{align}
\noindent 
where $w\left(n\right)$ represents additive environmental and system noises and $y\left(n\right)$ is a  reverberant echo that is nonlinearly generated from  $x\left(n\right)$. Before applying RES, the NLAEC system introduced in \cite{Ivry2021Nonlinear} is applied to reduce nonlinear echo. The NLAEC receives $m\left(n\right)$ as input and $x\left(n\right)$ as reference, and generates two signals: the echo estimate $\widehat{y}\left(n\right)$, and the desired-speech estimate $e\left(n\right)$, given by
\begin{align}
e\left(n\right) = m\left(n\right) - \widehat{y}\left(n\right) = s\left(n\right) + \left[y\left(n\right)-\widehat{y}\left(n\right)\right] + w\left(n\right).
\label{eq:errorModel}
\end{align}
\noindent The goal of the RES system is to suppress the residual echo \mbox{$y\left(n\right)-\widehat{y}\left(n\right)$} without distorting the desired-speech signal $s\left(n\right)$. 

\begin{figure}[t]
	\centering
	\includegraphics[width=\linewidth]{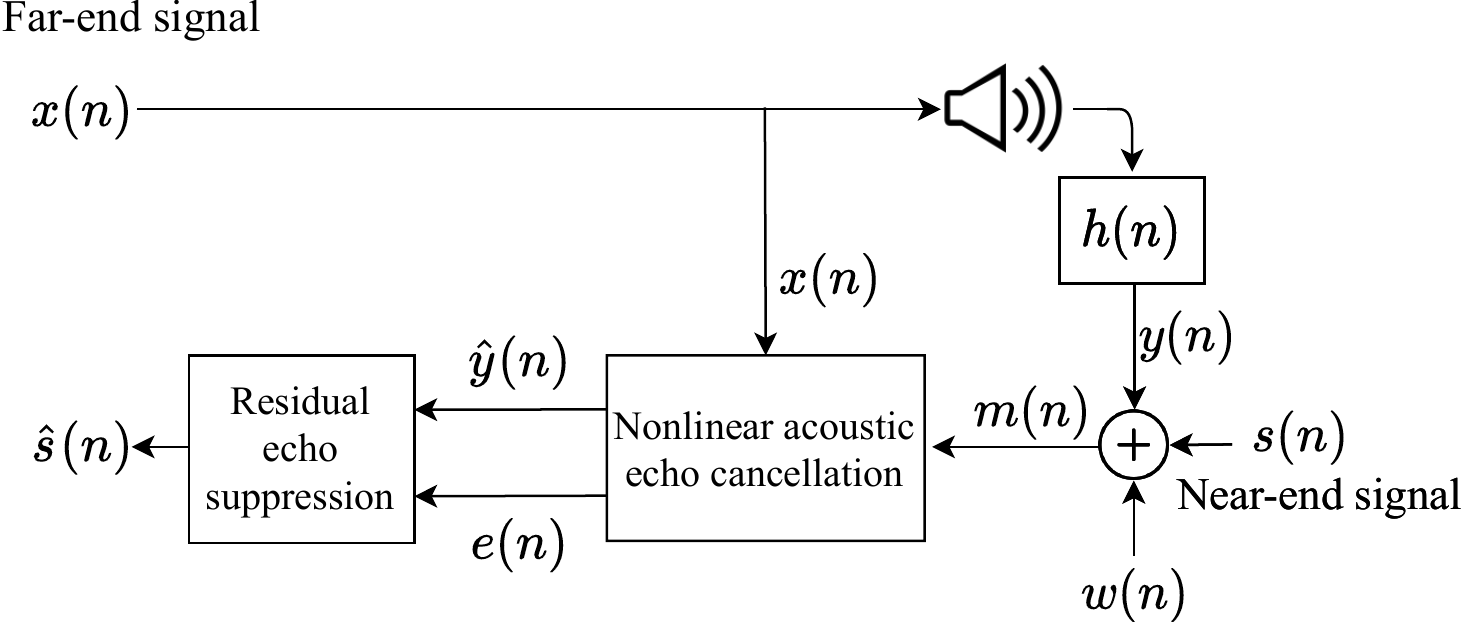}
	\caption{Residual-echo suppression scenario.}
	\label{fig:setup}
\end{figure}

\section{DSML and RESL}\label{sec:methods}
To derive the DSML and RESL, a deep learning-based RES system is considered as a time-varying gain. During double-talk, \mbox{$e\left(n\right)\neq 0$} and the gain is given by 
\begin{align}
g\left(n\right) = \frac{\widehat{s}\left(n\right)}{e\left(n\right)}\bigg|_{\textrm{Double-talk}}\textrm{ .}
\label{eq:DSML}
\end{align}Before introducing the DSML and RESL metrics, the SDR and its drawbacks are examined. According to \cite{vincent2006performance}, the SDR is defined as
\begin{equation}
\begin{aligned}
    \textrm{SDR} & = 10\log_{10}\frac{\Vert s\left(n\right)\Vert_{2}^{2}}{\Vert s\left(n\right)-\widehat{s}\left(n\right)\Vert_{2}^{2}} \bigg|_{\textrm{Double-talk}} \\ 
    & = 10\log_{10}\frac{\Vert s\left(n\right)\Vert_{2}^{2}}{\Vert s\left(n\right)-g\left(n\right) e\left(n\right)\Vert_{2}^{2}} \bigg|_{\textrm{Double-talk}}.
\label{eq:sdr}
\end{aligned}
\end{equation}
The SDR is affected by both the desired-speech distortion and residual-echo presence, and makes no distinction between cases in which \mbox{$g\left(n\right) e\left(n\right)$} comprises distortion-free speech and echo, or distorted speech without echo. Thus, the SDR does not correlate well with human ratings \cite{reddy2020dnsmos}, since these scenarios clearly exhibit different human perception ratings and different DNSMOS values. A distinction between desired-speech distortion and residual-echo suppression is extremely valuable for evaluating RES during double-talk. Hence, we propose two objective metrics by applying $g\left(n\right)$ separately to the desired speech and noisy residual-echo estimate. 

Formally, the DSML is calculated similarly to the SDR, but $g\left(n\right)$ is applied only to the desired speech $s\left(n\right)$:
\begin{align}
\textrm{DSML} = 10\log_{10}\frac{\Vert \tilde{s}\left(n\right)\Vert_{2}^{2}}{\Vert \tilde{s}\left(n\right) - g\left(n\right) s\left(n\right)\Vert_{2}^{2}} \bigg|_{\textrm{Double-talk}}.
\label{eq:resl}
\end{align}
\noindent The RESL is derived by estimating the noisy residual-echo as {$r\left(n\right) = e\left(n\right) - s\left(n\right)$}, and evaluating the following ratio:
\begin{align}
\textrm{RESL} = 10\log_{10}\frac{\Vert r\left(n\right)\Vert_{2}^{2}}{\Vert g\left(n\right) r\left(n\right)\Vert_{2}^{2}} \bigg|_{\textrm{Double-talk}} .
\end{align}
Note that the RES system may introduce a constant attenuation that leads to an artificial desired-speech distortion in the DSML. To ensure the DSML is invariant to that attenuation, it is compensated as in \cite{carbajal2018multiple}. Explicitly, $\tilde{s}\left(n\right) = \widehat{g}\left(n\right)s\left(n\right)$, where:
\begin{align}
\widehat{g}\left(n\right)=\frac{\bigl <g\left(n\right) s\left(n\right), s\left(n\right)\bigr>}{\Vert s\left(n\right)\Vert_{2}^{2}} .
\end{align}
% Formally, the DSML is calculated similarly to the SDR, but $g\left(n\right)$ is applied only to the desired speech $s\left(n\right)$:
% \begin{align}
% \textrm{DSML} = 10\log_{10}\frac{\Vert s\left(n\right)\Vert_{2}^{2}}{\Vert s\left(n\right) - g\left(n\right) s\left(n\right)\Vert_{2}^{2}} \bigg|_{\textrm{Double-talk}}.
% \label{eq:resl}
% \end{align}
% \noindent The RESL is derived by estimating the noisy residual-echo as {$r\left(n\right) = e\left(n\right) - s\left(n\right)$}, and evaluating the following ratio:
% \begin{align}
% \textrm{RESL} = 10\log_{10}\frac{\Vert r\left(n\right)\Vert_{2}^{2}}{\Vert g\left(n\right) r\left(n\right)\Vert_{2}^{2}} \bigg|_{\textrm{Double-talk}} .
% \end{align}
% Note that the RES system may introduce a constant attenuation that leads to both an artificial desired-speech distortion in the DSML and an artificial residual-echo suppression in the RESL. To ensure that the DSML and RESL are invariant to that attenuation, they are both compensated as shown in \cite{carbajal2018multiple}.  Also, higher DSML and 
% RESL values represent better performance, as
% higher DSML indicates less speech distortion and higher RESL  indicates stronger residual-echo suppression.

\section{RES System with a Design Parameter} \label{sec:system}
To evaluate the performances of the DSML and RESL, we employ a deep learning-based RES system that embeds a tunable design parameter \cite{Ivry2021Deep}. This system comprises a UNet neural network \cite{ronneberger2015u} with two input channels and one output channel. The network is fed with the short-time Fourier transform (STFT) \cite{griffin1984signal} amplitude of the NLAEC outputs and aims to recover the STFT amplitude of the desired speech. The design parameter $\alpha\geq0$ is embedded in a custom loss function $J(\alpha)$ that is minimized during training:
\begin{align}
J(\alpha) =
\Vert \widehat{S}\left(f\right)-S\left(f\right) \Vert^{2}_{2} + \alpha\Vert \widehat{S}\left(f\right) \Vert^{2}_{2}+ 0.1\,\sigma^{2}_{\widehat{S}\left(f\right) }\mathbb{I}_{\alpha> 0} \textrm{ ,}
\label{eq: loss}
\end{align}
\noindent where $\widehat{S}\left(f\right)$ and $S\left(f\right)$, respectively, represent the desired-speech prediction and ground truth spectra amplitudes, $\sigma^{2}_{\widehat{S}\left(f\right)}$ denotes the variance of $\widehat{S}\left(f\right)$, and $\mathbb{I}_{\alpha> 0}$ equals 1 when ${\alpha>0}$ and 0 otherwise. During the training stage, $J(\alpha)$ is minimized while $\alpha$ penalizes $\Vert \widehat{S}\left(f\right) \Vert^{2}_{2}$, which allows a dynamic tradeoff between the desired-speech distortion and residual-echo suppression of the system, namely between the DSML and RESL. When $\alpha=0$, the error between the desired-speech prediction and ground truth is minimized. However, when $\alpha>0$, smaller prediction values are generated. This reduces the level of residual echo but compromises the level of desired-speech distortion. $\sigma^{2}_{\widehat{S}\left(f\right)}$ mitigates sub-band nullification that may occur when $\alpha>0$. Note that $\alpha$ and the DSML-RESL tradeoff it creates can be tuned during the training process.

\section{Experimental Setup}\label{sec:setup}
\subsection{Database}\label{sec:database}
Two data corpora were employed in this study; the AEC-challenge database \cite{Culter2021Interspeech}, and a database recorded in our lab, both sampled at $16$~kHz. These corpora consider single-talk and double-talk periods both without and with echo-path change. In the former there is no movement during the recording, and in the latter either the near-end speaker or device are moving during the recording. In \cite{Culter2021Interspeech}, two open sources of synthetic and real recordings are introduced. The synthetic data includes $100$~h, and the real data contains $140$~h of audio clips, generated from $5,000$ hands-free devices that are used in various acoustic environments. In both real and synthetic cases, signal-to-echo ratio (SER) and signal-to-noise ratio (SNR) levels were distributed on $\left[-10,10\right]$ dB and $\left[0,40\right]$ dB, respectively. 
Additional real recordings were conducted in our lab to test the generalization of the DSML and RESL to unseen setups and their robustness to extremely low levels of SERs. This database is fully described in \cite{Ivry2021Deep}. For completion, it contains $40$~h of recordings from the TIMIT \cite{TIMIT_correct} and LibriSpeech \cite{panayotov2015librispeech} corpora with SNR levels of $32\pm5$ dB and SER levels distributed on $\left[-20,-10\right]$ dB. 

The SER is defined as \mbox{SER=$10\log_{10}\left[\Vert s\left(n\right) \Vert_{2}^{2} / \Vert y\left(n\right) \Vert_{2}^{2}\right]$} and the SNR is defined as \mbox{SNR=$10\log_{10}\left[\Vert s\left(n\right) \Vert_{2}^{2} / \Vert w\left(n\right) \Vert_{2}^{2}\right]$} in dB, each is calculated with 50\% overlapping time frames of 20 ms.

\subsection{Data Processing, Training, and Testing}\label{preproc}
The real and synthetic data from \cite{Culter2021Interspeech} was randomly split to create 185~h of training set and 45~h of validation set. The test set contains only real data that includes the remaining 10~h from \cite{Culter2021Interspeech} and all 40~h from \cite{Ivry2021Deep}. Each set was divided into 10~s segments that contain recordings in different setups. This leads to frequent re-convergence during transitions between segments, both with and without echo-path change. These sets are balanced to prevent bias in the results, as detailed in \cite{Ivry2021Deep}. The NLAEC system, which is also deep learning-based \cite{Ivry2021Nonlinear}, and the succeeding RES system \cite{Ivry2021Deep}, were trained separately. During testing, in accordance with Section \ref{sec:methods}, the artificial gain that may be introduced by the RES system is compensated as in \cite{vincent2006performance, carbajal2018multiple} before deriving the DSML and RESL.

\subsection{Additional Performance Metrics}
We employ additional metrics to evaluate RES. The echo return loss enhancement (ERLE) \cite{ERLE} measures echo reduction between the degraded and enhanced signals when only echo and noise are present:
\begin{align}
\textrm{ERLE} =
10\log_{10}\frac{\Vert e\left(n\right)\Vert_{2}^{2}}{\Vert \widehat{s}\left(n\right)\Vert_{2}^{2}} \bigg|_{\textrm{Far-end single-talk}} \textrm{ .}
\label{eq: erle}
\end{align}
The signal-to-artifacts ratio (SAR) \cite{vincent2006performance} measures the desired-speech distortion during near-end single-talk periods:
\begin{align}
\textrm{SAR} =
10\log_{10}\frac{\Vert s\left(n\right)\Vert_{2}^{2}}{\Vert s\left(n\right)-\widehat{s}\left(n\right)\Vert_{2}^{2}} \bigg|_{\textrm{Near-end single-talk}} \textrm{ .}
\label{eq: sar}
\end{align}
The perceptual evaluation of speech quality (PESQ) \cite{PESQ} metric, which correlates well with the DNSMOS \cite{reddy2020dnsmos}, is used in double-talk. The SAR and SDR are compensated as in Section \ref{sec:methods}.
% We employ additional metrics that are widely used to evaluate RES systems. The echo return loss enhancement (ERLE) \cite{ERLE} measures echo reduction between the degraded and enhanced signals when only echo and noise are present
% \begin{align}
% \textrm{ERLE} =
% 10\log_{10}\frac{\Vert e\left(n\right)\Vert_{2}^{2}}{\Vert \widehat{s}\left(n\right)\Vert_{2}^{2}} \Bigr|_{\textrm{Far-end single-talk}} \textrm{ .}
% \label{eq: erle}
% \end{align}
% The signal-to-artifacts ratio (SAR) \cite{vincent2006performance} measures the desired-speech distortion during near-end single-talk periods 
% \begin{align}
% \textrm{SAR} =
% 10\log_{10}\frac{\Vert s\left(n\right)\Vert_{2}^{2}}{\Vert s\left(n\right)-\widehat{s}\left(n\right)\Vert_{2}^{2}} \Bigr|_{\textrm{Near-end single-talk}} \textrm{ .}
% \label{eq: sar}
% \end{align}
% Beside the DNSMOS, SDR, DSML, and RESL, double-talk periods are also evaluated with the perceptual evaluation of speech quality (PESQ) \cite{PESQ} metric that highly correlates with the DNSMOS \cite{reddy2020dnsmos}. 

\begin{figure}[t!]
	\centering
	\begin{subfigure}[t]{0.232\textwidth}
		\centering
		\includegraphics[width=\textwidth]{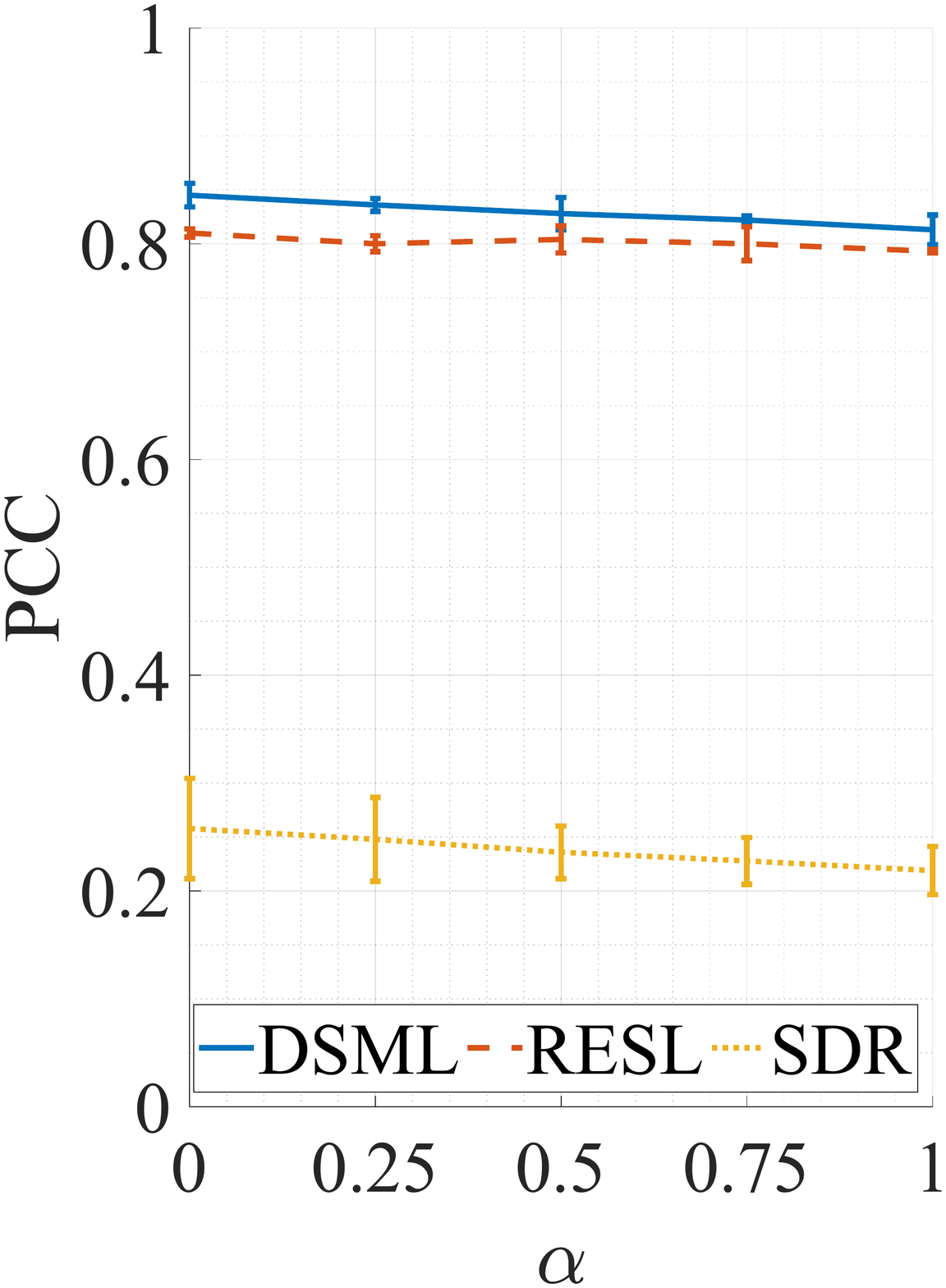}
	\end{subfigure}
	\hfill
	\begin{subfigure}[t]{0.232\textwidth}
		\centering
		\includegraphics[width=\textwidth]{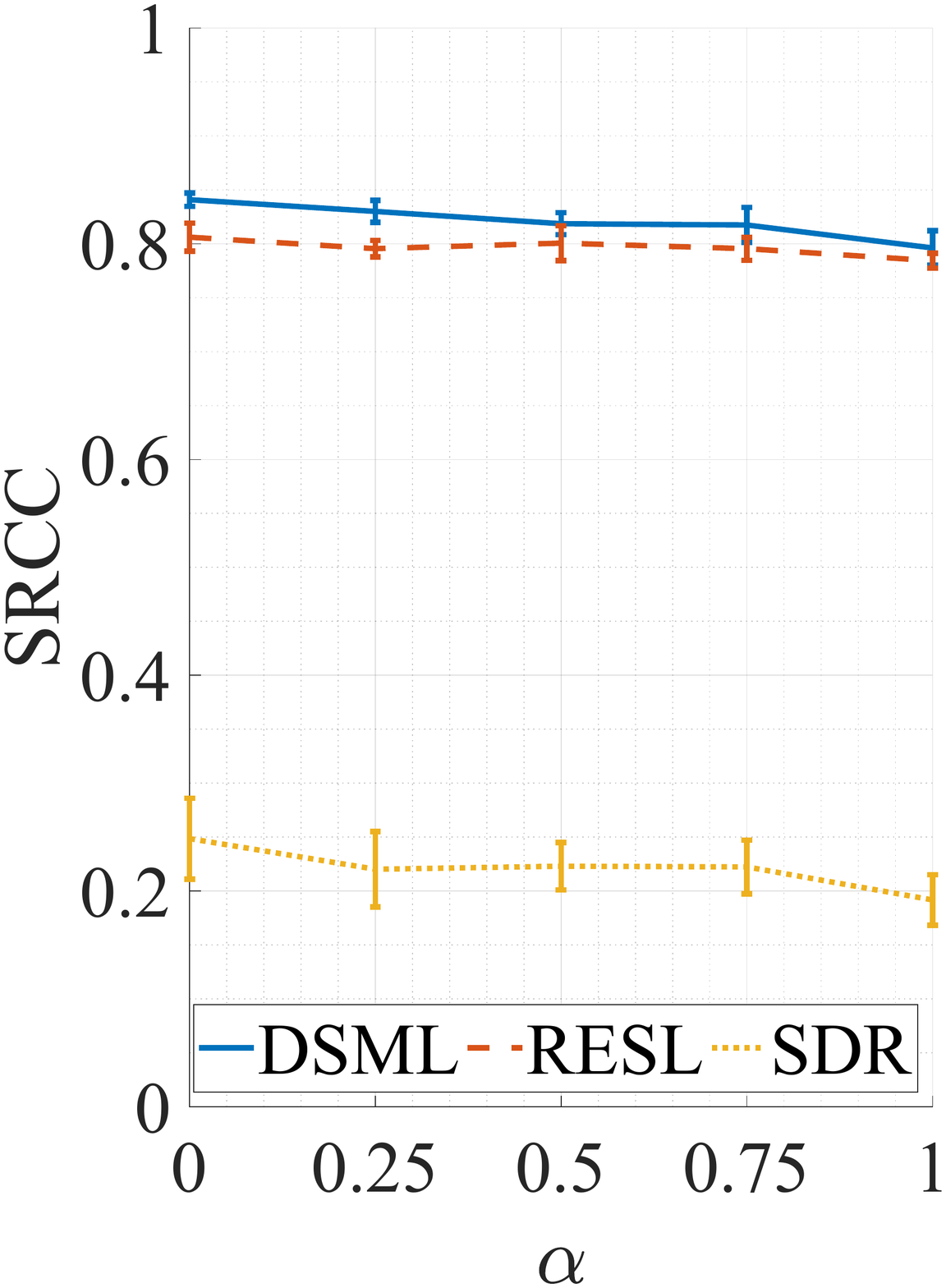}
	\end{subfigure}
\caption{Correlation of DNSMOS with the proposed DSML and RESL metrics, and the widely-used SDR.}
\label{fig:corr_scores}
\end{figure}
\begin{figure}[t]
	\centering
	\begin{subfigure}[t]{0.232\textwidth}
		\centering
		\includegraphics[width=\textwidth]{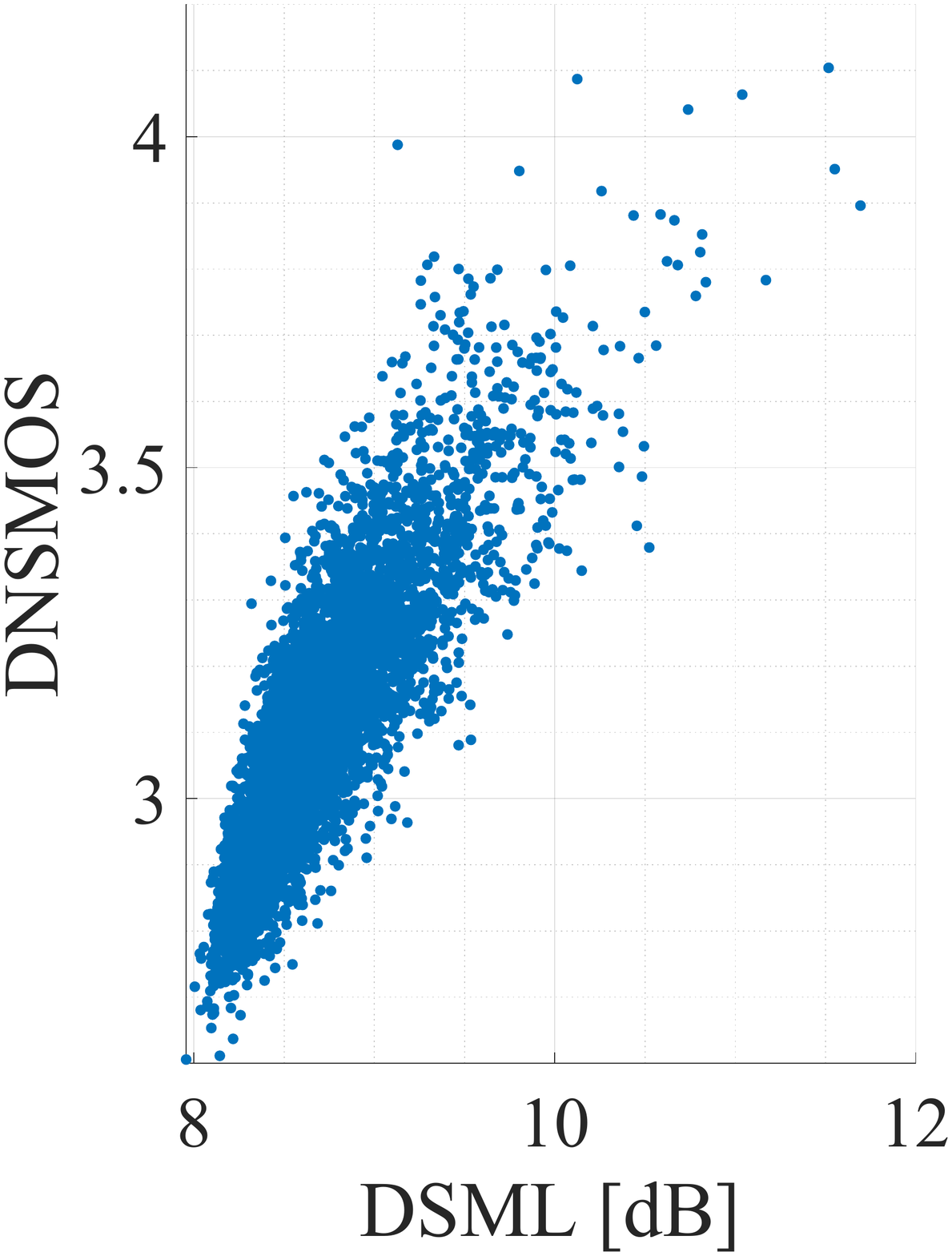}
		% \caption{DNSMOS vs DSML}
	\end{subfigure}
	~
	\begin{subfigure}[t]{0.232\textwidth}
		\centering
		\includegraphics[width=\textwidth]{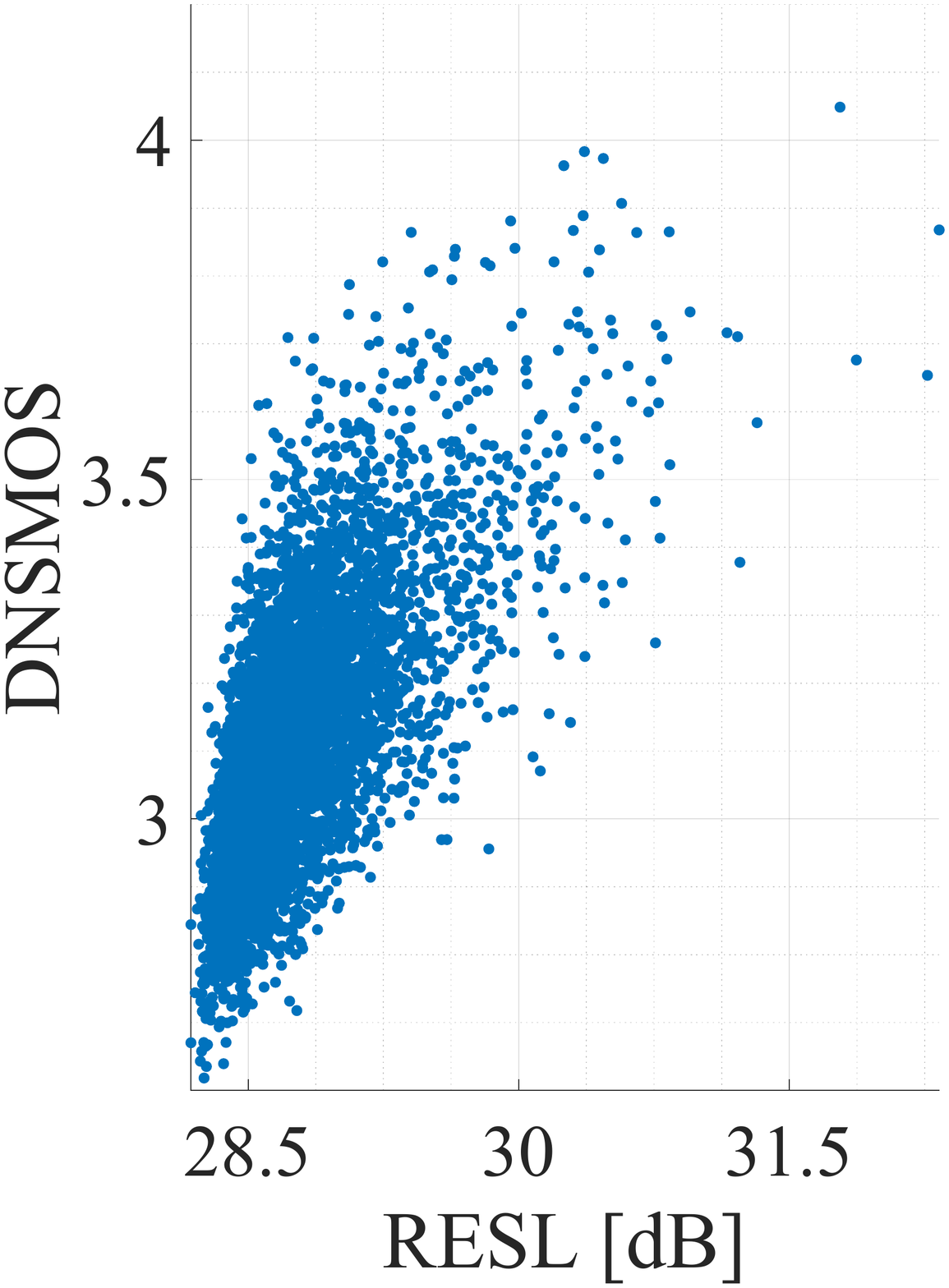}
		% \caption{DNSMOS vs RESL}
	\end{subfigure}
    ~
	\begin{subfigure}[t]{0.232\textwidth}
		\centering
		\includegraphics[width=\textwidth]{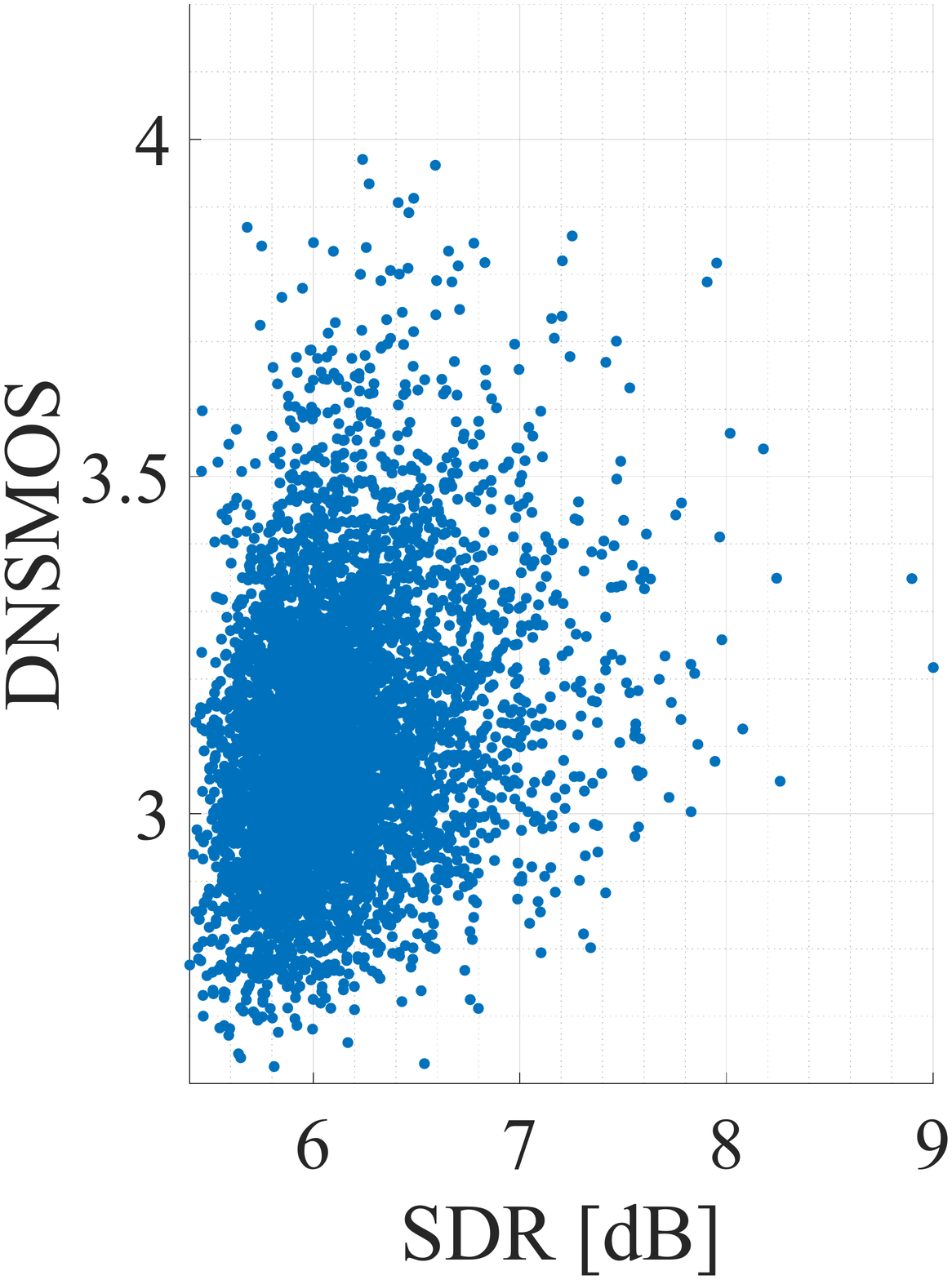}
		% \caption{DNSMOS vs SDR}
	\end{subfigure}
\caption{Scatter plots of DNSMOS versus the proposed DSML and RESL metrics, and the widely-used SDR.}
\label{fig:corr_scatter}
\end{figure}

\section{Experimental Results}\label{sec:results}
The performance metrics are evaluated using the RES system and are calculated with 50\% overlapping frames of 20~ms. The metrics are reported by their mean and standard deviation (std) values in Table \ref{table:perf_table}, and by their mean only in Figures \ref{fig:tradeoff} and \ref{fig:snr_ser}, with respect to the test set specified in each experiment. For all metrics, higher mean and lower std indicate a better performance. In our study, the convergence of the NLAEC follows the definitions in \cite{Ivry2021Nonlinear,paleologu2015overview}, and the DNSMOS is calculated using the API provided by Microsoft \cite{reddy2020dnsmos}. 

First, we explore the correlation of the DSML and RESL with the DNSMOS using Pearson correlation coefficient (PCC) \cite{benesty2009pearson} and Spearman's rank correlation coefficient (SRCC) \cite{gauthier2001detecting}, as done in \cite{reddy2020dnsmos,sridhar2020icassp}. This experiment includes segments without echo-path change after NLAEC convergence for \mbox{$\alpha=\left[0,0.25,0.5,0.75,1\right]$}, and the results are shown in Figure \ref{fig:corr_scores}. The conclusion drawn in \cite{reddy2020dnsmos} is reaffirmed in this study, i.e., the SDR does not correlate well with the DNSMOS, as the PCC and SRCC mean values are below $0.26$ for all $\alpha$. On the contrary, the DSML and RESL are highly correlated with the DNSMOS, with mean correlation scores between $0.78$ and $0.85$ for all $\alpha$. Also, compared to the SDR, the DSML and RESL correlations are relatively more consistent across $\alpha$ values, as inferred from their lower std values. To visualize these correlations, Figure \ref{fig:corr_scatter} depicts scatter plots of the DNSMOS versus the DSML, RESL, and SDR metrics for random sample values with \mbox{$\alpha=0$}. These plots validate the poor correlation between the DNSMOS and SDR, and the high correlation between the DNSMOS and the DSML and RESL. Conclusively, the DSML and RESL are better correlated with human perception and speech quality evaluation.

All performance metrics are evaluated in Table~\ref{table:perf_table} with $\alpha=0$. Separate results are shown for segments without and with echo-path change after NLAEC convergence, and for segments before convergence. The DSML and RESL are consistent with all other metrics, which degrade when shifting from no echo-path change to echo-path change scenarios, and further degrade when considering segments before convergence. This also implies high generalization of the DSML and RESL to various setups. The DSML is consistently higher than the SDR, as expected, since the definition in (\ref{eq:sdr}) also considers echo and noise in the denominator. Also, the DSML is lower than the SAR, which is applicable to single-talk segments where speech is less distorted by the RES system. The RESL is always lower than the ERLE, which is relevant to segments without desired speech where echo is more suppressed. These observations highlight the reliability of the DSML and RESL metrics.

Next, the relation between tuning $\alpha$ and the DSML-RESL tradeoff it creates is investigated. Figure \ref{fig:tradeoff} considers segments without and with echo-path change after convergence, and segments before convergence, for \mbox{$\alpha=\left[0,0.25,0.5,0.75,1\right]$}. As $\alpha$ increases, speech is more distorted and the DSML decreases, while residual echo is more suppressed and the RESL increases. This tradeoff occurs across all scenarios and is empirically consistent for all $\alpha$ values. This tradeoff is also analyzed in various SER and SNR levels that occur in real-life setups. In this experiment, segments without echo-path change are considered and results are given in Figure~\ref{fig:snr_ser}. It can be observed that both the DSML and RESL are impaired when acoustic conditions deteriorate, as expected. Also, the relation between $\alpha$ and the metrics is retained, i.e., for all levels of echo and noise, increasing $\alpha$ degrades the DSML and enhances the RESL.

Finally, we offer a practical design scheme for possible dynamic user requirements. Assume an environment without echo-path change after convergence, which can be inferred by the user using the definitions in \cite{Ivry2021Nonlinear,paleologu2015overview}. At first, the user requires an average RESL higher than $30$ dB and DSML higher than $8.4$ dB. According to Figure \ref{fig:tradeoff}(a), $\alpha=0.5$ is selected. Next, the user evaluates that $\textrm{SER}=0$ dB and $\textrm{SNR}=20$ dB, e.g., by respectively analyzing double-talk and near-end single-talk periods, and accordingly decides to suppress the maximal amount of echo that maintains DSML no lower than $8.3$ dB. Then, according to Figure~\ref{fig:snr_ser}, the user shifts $\alpha=0.5$ to $\alpha=0.75$ during training, which decreases the average DSML to $8.3$ dB and increases the average RESL to above $31$ dB.

\begin{table}[]
    \renewcommand{\arraystretch}{1.5}
	\caption{Performance metrics in various scenarios with $\alpha=0$.}
	\label{table:perf_table}
	\centering
	\leavevmode
	\begin{tabularx}{0.45\textwidth}
	{
	>{\raggedright\arraybackslash \hsize=\hsize}Y Y Y Y
	}
    \heading{}
    & \heading{\vtop{\hbox{\strut \textrm{No echo-path}}\hbox{\strut \hspace{0.35cm} change}}} 
    & \heading{\vtop{\hbox{\strut \textrm{Echo-path}}\hbox{\strut \hspace{0.15cm} change}}}
    & \heading{\vtop{\hbox{\strut \textrm{\hspace{0.3cm}Before}}\hbox{\strut convergence}}}
    \\
    \hline
    
    DNSMOS & \hspace{0.15cm}{3.12 $\pm$ 0.2} & 2.91 $\pm$ 0.3 & 2.56 $\pm$ 0.6\\ \hline
    
    DSML & \hspace{0.15cm}{8.73 $\pm$ 0.4} & 8.34 $\pm$ 0.5 & 6.97 $\pm$ 0.7\\ \hline
    
    RESL & \hspace{0.15cm}{29.1 $\pm$ 3.7} & 25.9 $\pm$ 4.4 & 22.1 $\pm$ 5.6\\ \hline
    
    SDR & \hspace{0.15cm}{6.13 $\pm$ 0.4} & 5.94 $\pm$ 0.6 & 5.57 $\pm$ 0.8\\ \hline
    
    PESQ & \hspace{0.15cm}{3.58 $\pm$ 0.2} & 3.35 $\pm$ 0.5 & 3.18 $\pm$ 0.6\\ \hline
    
    SAR & \hspace{0.15cm}9.88 $\pm$ 0.4 & 9.69 $\pm$ 0.5 & 9.51 $\pm$ 0.6\\ \hline
    
    ERLE & \hspace{0.15cm}33.2 $\pm$ 3.1 & 29.1 $\pm$ 4.2 & 26.4 $\pm$ 5.1\\ \hline
	\end{tabularx}
\end{table}

\begin{figure}
	\centering
	\begin{subfigure}[t]{0.2\textwidth}
		\centering
		\includegraphics[width=\textwidth]{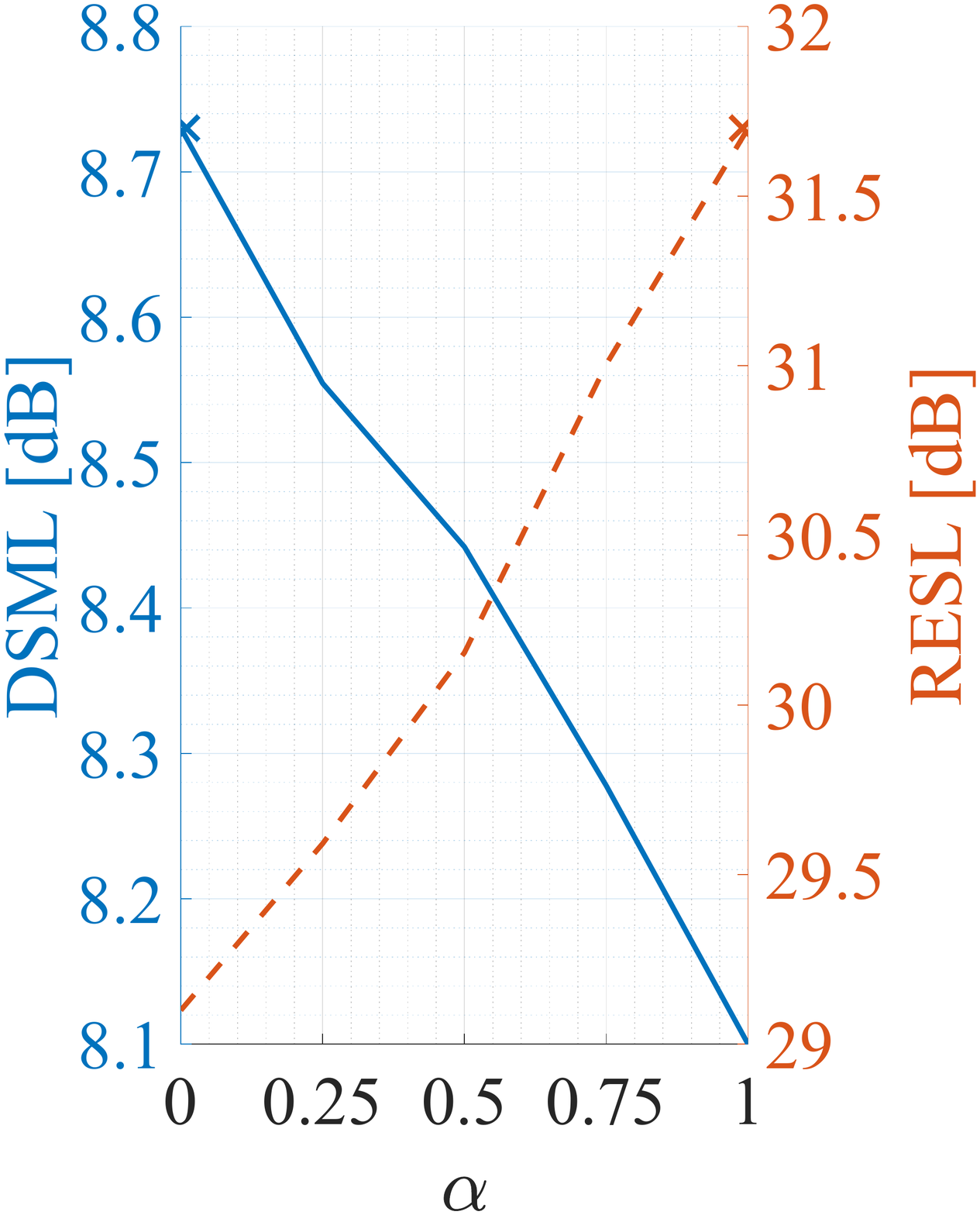}
		\caption{No echo-path change}
	\end{subfigure}
`   ~
	\begin{subfigure}[t]{0.2\textwidth}
		\centering
		\includegraphics[width=\textwidth]{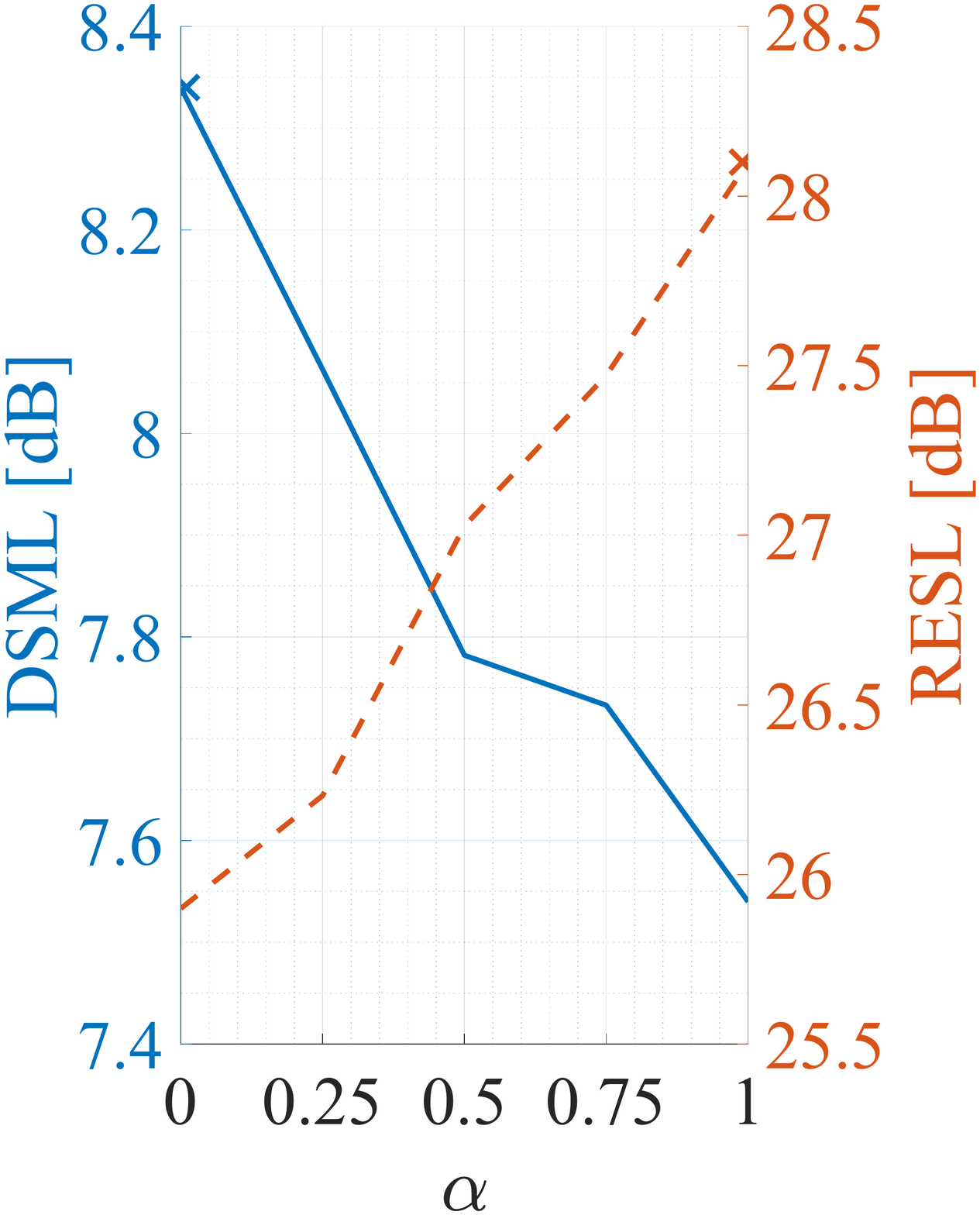}
		\caption{Echo-path change}
	\end{subfigure}
	~
	\begin{subfigure}[t]{0.2\textwidth}
		\centering
		\includegraphics[width=\textwidth]{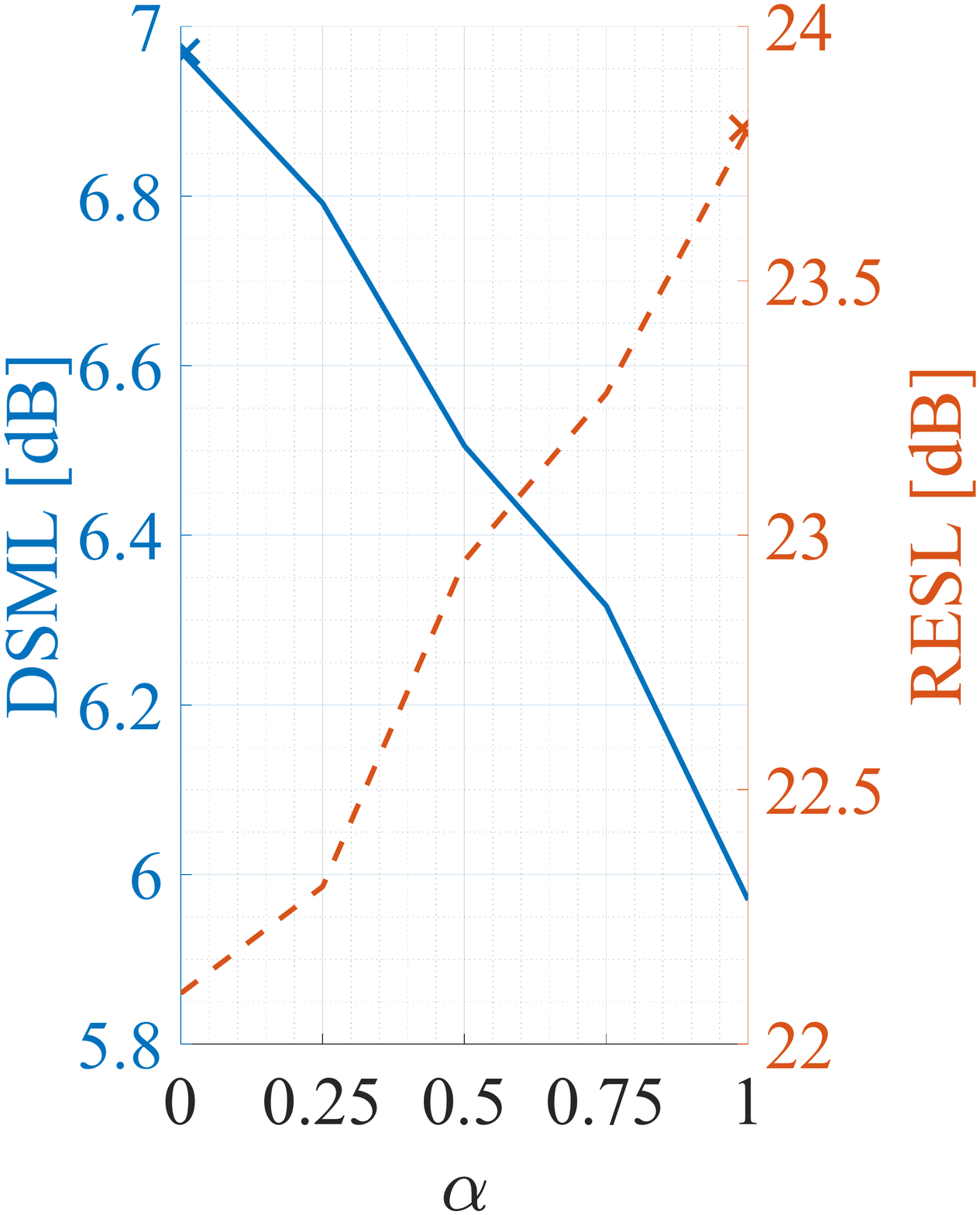}
		\caption{Before convergence}
	\end{subfigure}

\caption{DSML-RESL tradeoff for various values of $\alpha$.}
\label{fig:tradeoff}
\end{figure}

\begin{figure}
	\centering
	\begin{subfigure}[t]{0.2\textwidth}
		\centering
		\includegraphics[width=\textwidth]{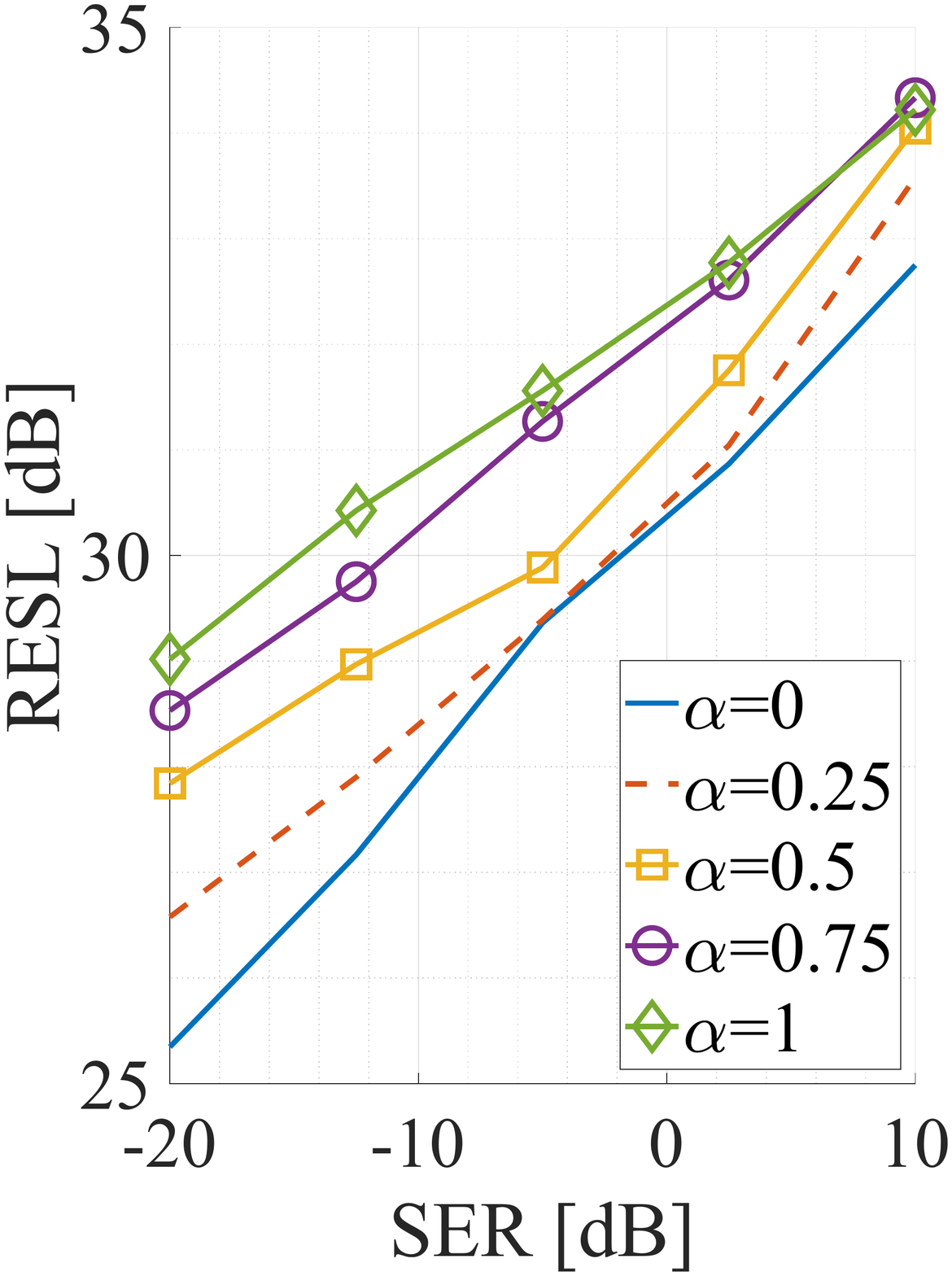}
	\end{subfigure}
`   ~
	\begin{subfigure}[t]{0.2\textwidth}
		\centering
		\includegraphics[width=\textwidth]{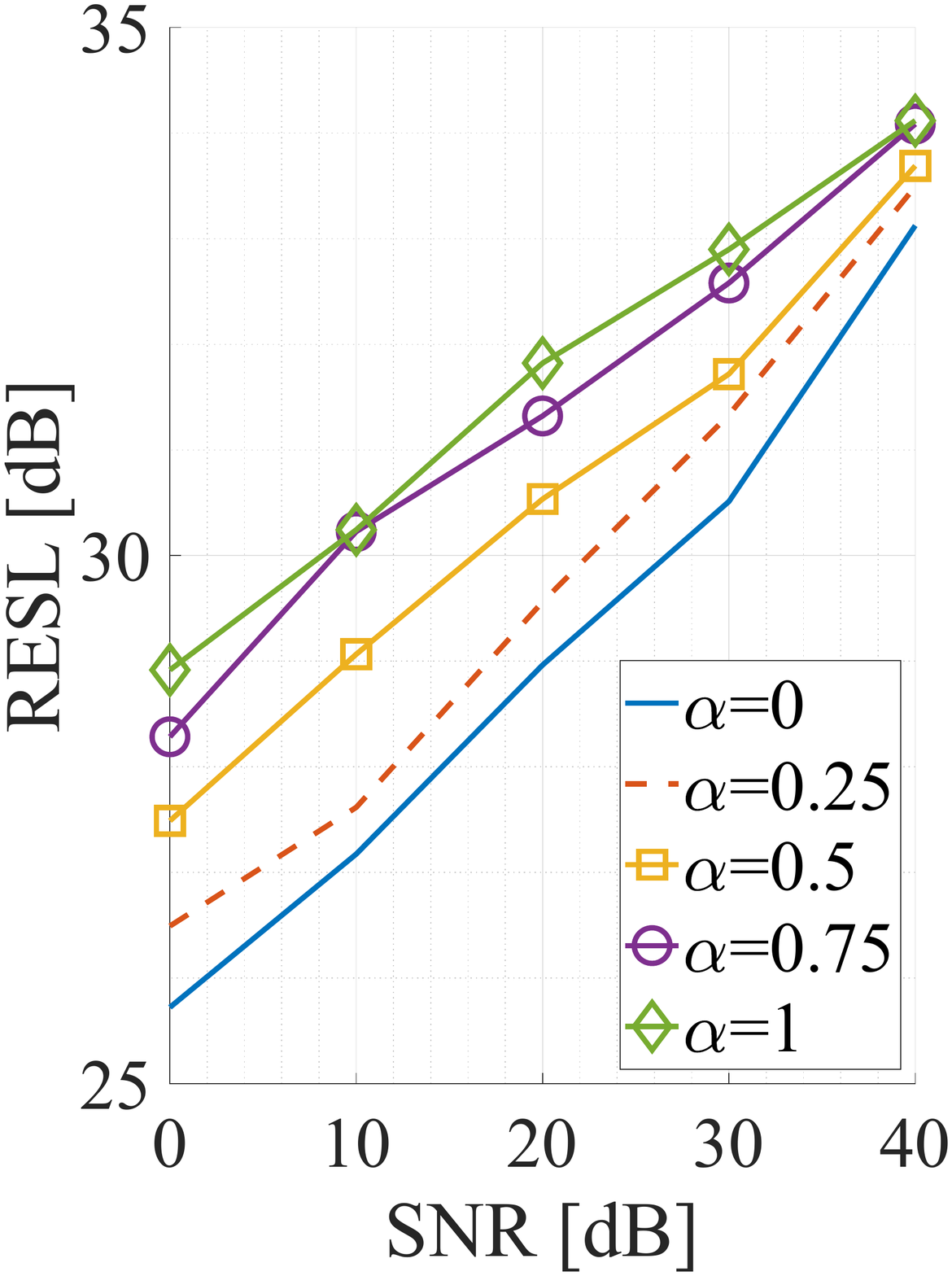}
	\end{subfigure}
	\hfill
	\begin{subfigure}[t]{0.2\textwidth}
		\centering
		\includegraphics[width=\textwidth]{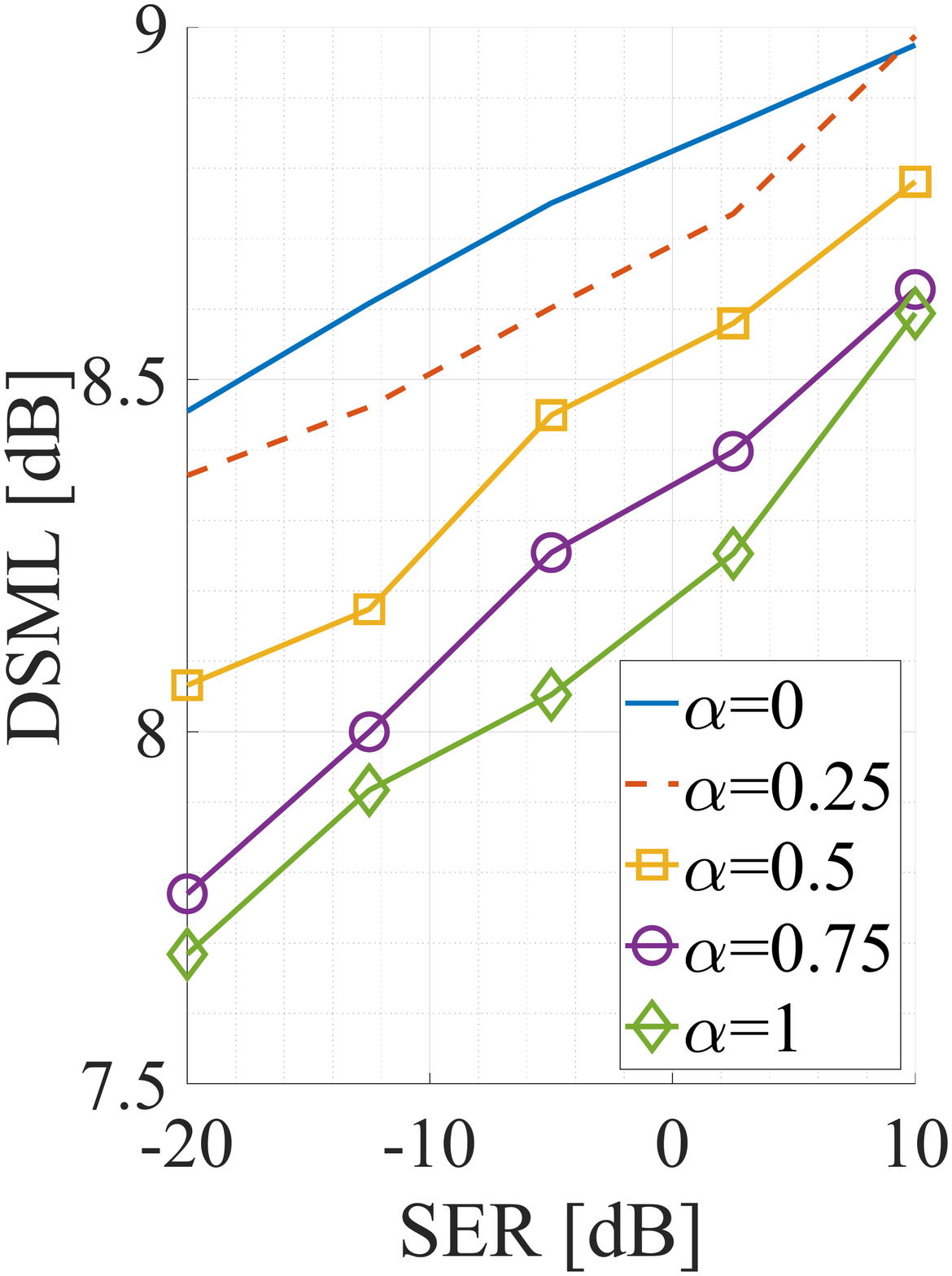}
	\end{subfigure}
    ~
    \begin{subfigure}[t]{0.2\textwidth}
		\centering
		\includegraphics[width=\textwidth]{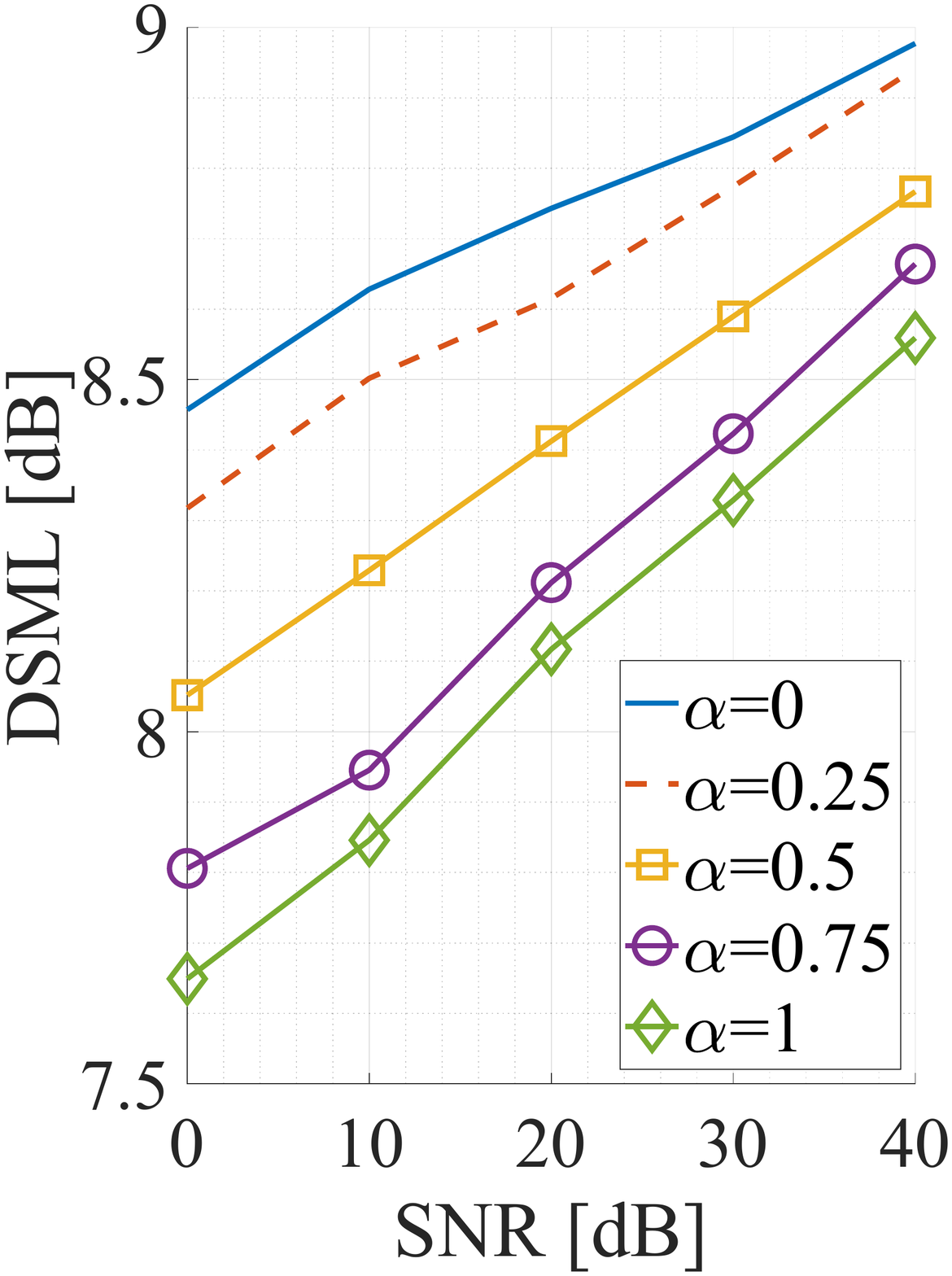}
	\end{subfigure}
\caption{DSML-RESL tradeoff for various values of $\alpha$ in different echo and noise levels.}
\label{fig:snr_ser}
\end{figure}

\section{Conclusion}\label{sec:conclusion}
We introduced two objective metrics to separately assess the desired-speech maintained level (DSML) and the residual-echo suppression level (RESL) during double-talk. The performances of these metrics are evaluated using a deep learning-based RES system with a tunable design parameter $\alpha$, with 280~h of real and synthetic recordings. We showed that the DSML and RESL correlate well with human perception compared to the popular SDR metric, which may suggest they are more suitable for speech quality evaluation. Also, we empirically learned the relation between tuning $\alpha$ and the resulting DSML-RESL tradeoff and offered a practical design scheme that benefits dynamic user preferences. Future work will analyze the DNSMOS as an appropriate evaluation for RES subjective quality in double-talk, and explore the DSML-RESL tradeoff to yield a practical design scheme for optimal speech quality.
% We have introduced two objective metrics to separately assess the desired-speech maintained level (DSML) and the residual-echo suppression level (RESL) during double-talk. The performances of these metrics are evaluated using a deep learning-based RES system with a tunable design parameter $\alpha$, using 280~h of real and synthetic recordings. We showed that the DSML and RESL correlate well with human perception compared to the widely-used SDR metric, and are thus more appropriate to assess speech quality during double-talk. Also, we empirically learned the relation between tuning $\alpha$ and the resulting DSML-RESL tradeoff and offered a practical design scheme that can benefit dynamic user preferences. 

\bibliographystyle{IEEEtran}
\bibliography{WASPAA_refs}

\end{sloppy}
\end{document}